\documentclass{PoS}

\title{Nuclear Parton Distributions at the future Electron-Ion Collider}

\ShortTitle{Nuclear Parton Distributions at the future Electron-Ion Collider}

\author{\speaker{Salvatore Fazio} \\
       Brookhaven National Laboratory\\
       E-mail: \email{sfazio@bnl.gov}}

\abstract{The 2015 nuclear physics long-range plan endorsed the realization of an Electron-Ion Collider (EIC) as the next large construction project after the completion of FRIB.  
With its high luminosity ( $> 10^{33} cm^{-2}s^{-1}$), wide kinematic reach in center-of-mass-energy (45~GeV to 145~GeV) and high lepton and proton beam polarization, an EIC provides an unprecedented opportunity to reach new frontiers in our understanding of the spin and dynamic structure of nuclei.
Despite of the success of the HERA collider in investigating the structure of a single nucleon, the partonic structure of nuclei at moderate-to-small Bjorken's $x$ still remains elusive. 
We present the evaluated impact of an EIC in extracting the nuclear structure-functions from measurements of the reduced cross section in deep inelastic scattering, including also the case of measuring heavy quark production events. The potential constraints offered by the EIC data in extracting the nuclear parton distribution functions is also discussed.}

\FullConference{ 25th International Workshop on Deep Inelastic Scattering and Related Topics \\
         3-7 April 2017 \\
         University of Birmingham, Birmingham, UK}

\begin{document}


Whilst the current knowledge of the parton distribution functions (PDFs) on a free nucleus largely derive from the e+p collider experiments at HERA, a precise determination of the partonic content of a bound nucleus is still an outstanding goal in nuclear physics. The realization of an Electron-Ion Collider (EIC) will be key for constraining the nuclear PDFs (nPDFs).  A precise knowledge of nPDFs is crucial in studying the transition between linear and non-linear scale evolution of the parton densities and a regime, known as ``saturation", where the recombination of gluons at low $x$ becomes increasingly important and the growth of the gluon density eventually tames. nPDFs are also essential for the theoretical interpretation of the A+A and p+A data from collider experiments at RHIC and the LHC. 
 
 \begin{figure}[t] 
   \centering
   \includegraphics[width=0.49\textwidth]{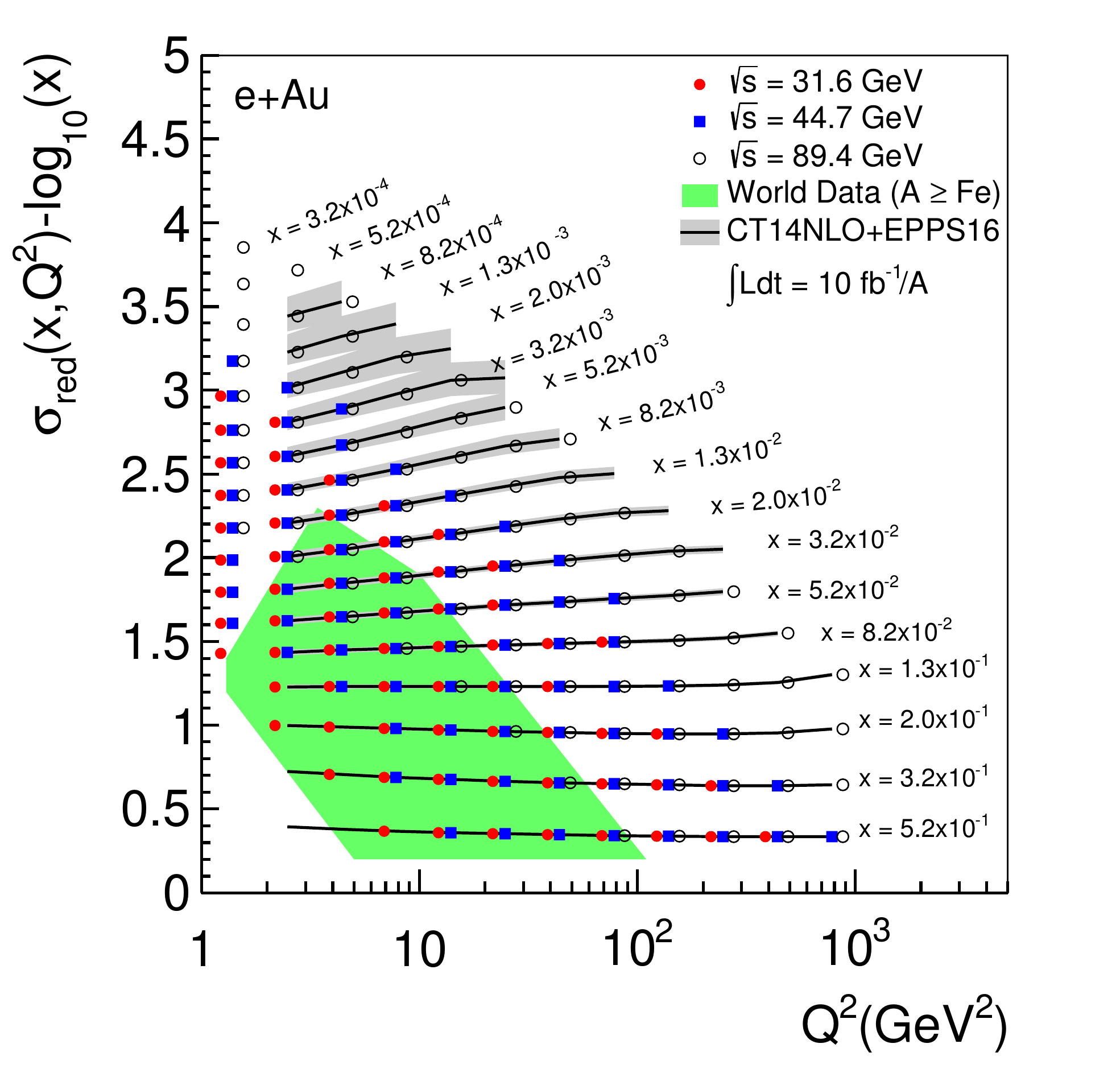}   
   \includegraphics[width=0.49\textwidth]{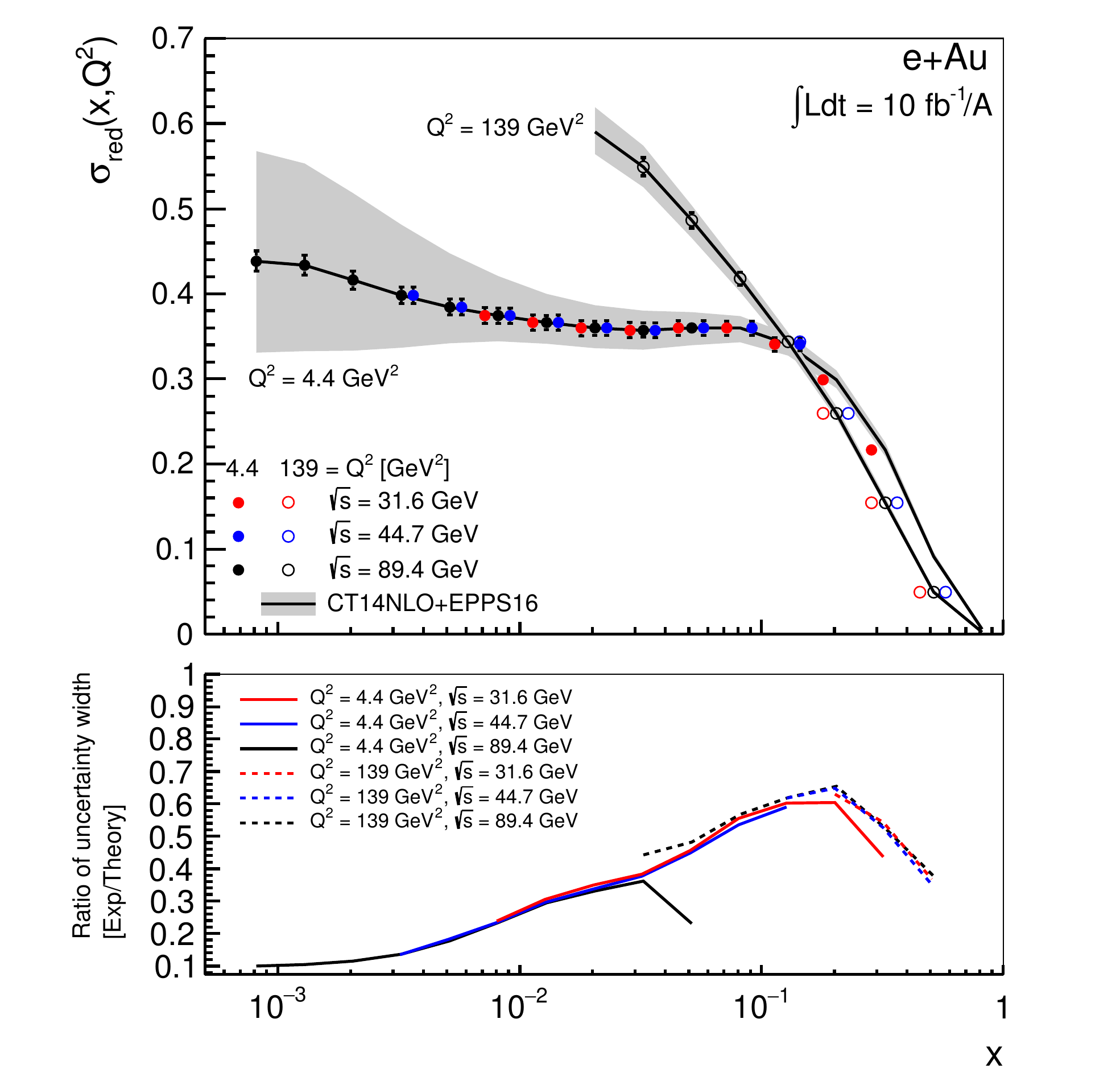}
   \caption{The reduced cross section ({\it left}) in $e$+Au collisions at an EIC is plotted as a 
function of $Q^2$ and $x$. The points are shifted by -log$_{10}(x)$ for visibility. 
Two examples of the $\sigma_{r}$ ({\it right}) at $Q^{2}$ values of 4.4~GeV$^{2}$ and 139~GeV$^{2}$  are plotted versus $x$, 
with the bottom panel showing the ratio between the widths of the experimental and theoretical uncertainties. Points that correspond to different energy 
configurations are horizontally offset in $Q^2$ for visibility. The vertical bars represent statistical and systematic uncertainties added in quadrature. The overall 1.4\% systematic uncertainty on the luminosity measurement is not considered in the plot.}
   \label{Fig:SigmaRed} 
\end{figure}
 
The DIS cross section, a direct observable used in constraining the nPDFs, is customarily divided by the Mott cross section and expressed in a dimensionless form known as the reduced cross section, $\sigma_{r}$, which in turn can be expressed in terms of the structure functions $F_{2}$ and $F_{L}$ 

\begin{equation}
\sigma_{r} = F_2(x,Q^2) - \frac{y^2}{1 + (1 - y)^2} F_L(x,Q^2).
\label{Eq:SigmaRed}
\end{equation}
The momentum distributions of (anti)quarks can be measured through scaling violation fits of $F_{2}$, whereas $F_{L}$ has a direct contribution from gluons~\cite{Armesto2011}. 
In addition, at an EIC it will be possible to obtain a direct constraint of the gluon density by measuring quark-gluon fusion processes like the charm production. This will also allow us to study different heavy quark mass schemes and constrain the intrinsic heavy-flavor components in the nPDFs~\cite{Accardi:2016ndt}. 

For the results presented here, e+Au collisions have been simulated using the PYTHIA 6.4~\cite{Pythia64} Monte Carlo (MC) generator augmented with the EPS09~\cite{EPS09} nuclear PDFs. Different beam-energy configurations have been considered, corresponding to a range in c.o.m. energy from 30 to 90 GeV. The assumed bin-by-bin systematic uncertainty on the measurements of inclusive DIS and charm production is 1.6\% and 3.5\% respectively. The additional overall systematic uncertainty on the measurement of luminosity is estimated to be 1.4\%.

Figure~\ref{Fig:SigmaRed} ({\it left}) shows $\sigma_{r}$ versus $Q^2$ at different $x$ values, for three c.o.m. energies. A log$_{10}(x)$ factor is subtracted and points at the same $x$ are shifted for visibility. Simulated data are compared to the latest theoretical predictions from the EPPS16~\cite{EPPS16} nuclear PDFs. 
For the purpose of better appreciating the experimental uncertainties, Figure~\ref{Fig:SigmaRed}~({\it right}) shows two examples of the $\sigma_{\rm r}$ as a function of $x$ at $Q^{2} = 4.4$ and 139~GeV$^{2}$ without the subtraction of the large log$_{10}(x)$ factor. In the bottom panel, the ratios between the full widths of the experimental and theoretical uncertainties are plotted versus $x$ for the different c.o.m energies. One can see that at small $x$, and small $Q^2$, the expected uncertainties on inclusive $\sigma_{r}$ measurements at an EIC are much smaller than those from current theoretical prediction. At larger $x$ values, the current constraints from fixed-target experiments (SLAC and NMC) are already stringent.

\begin{figure}[t] 
   \centering
      \includegraphics[width=0.49\textwidth]{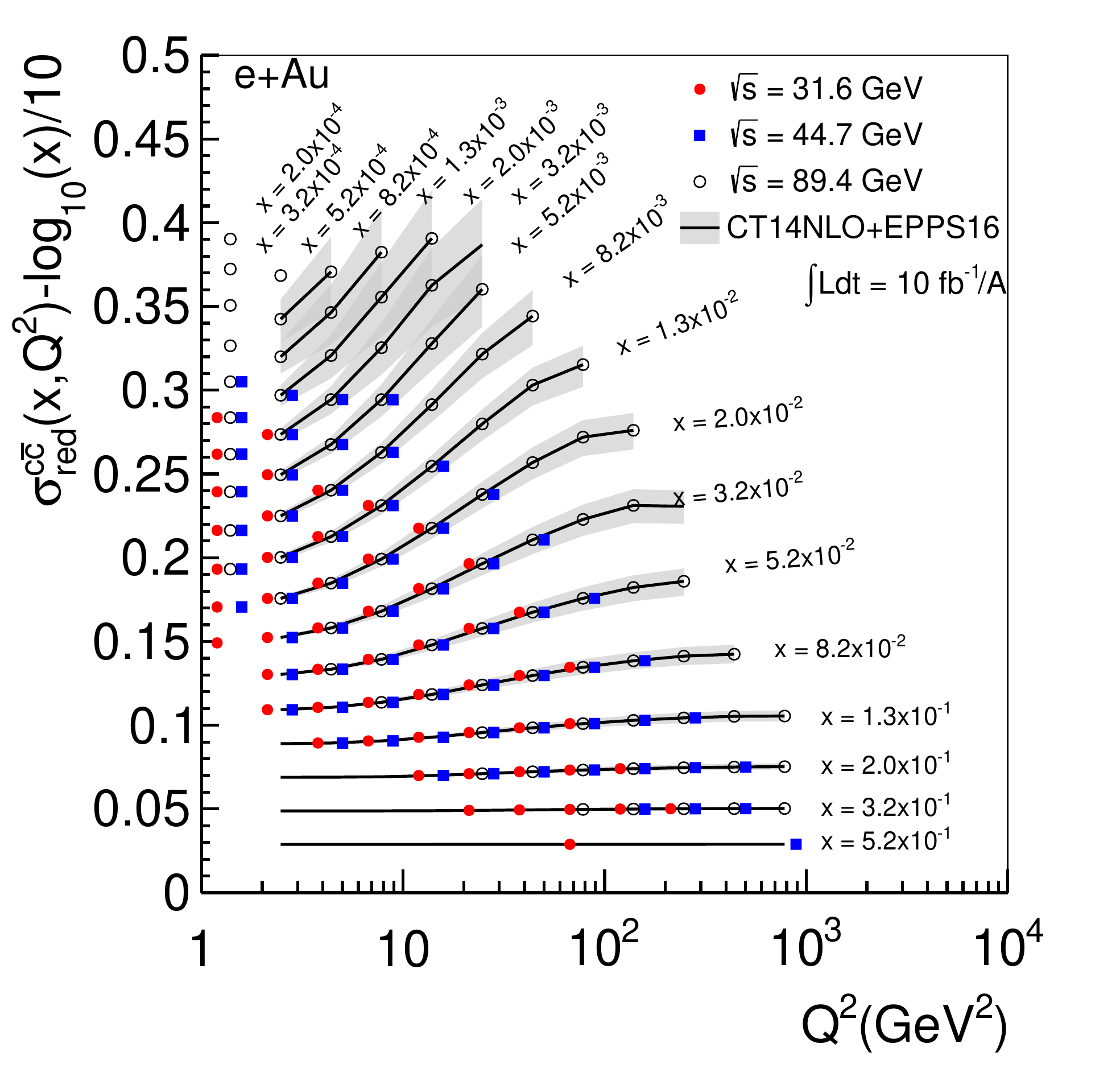}   
      \includegraphics[width=0.49\textwidth]{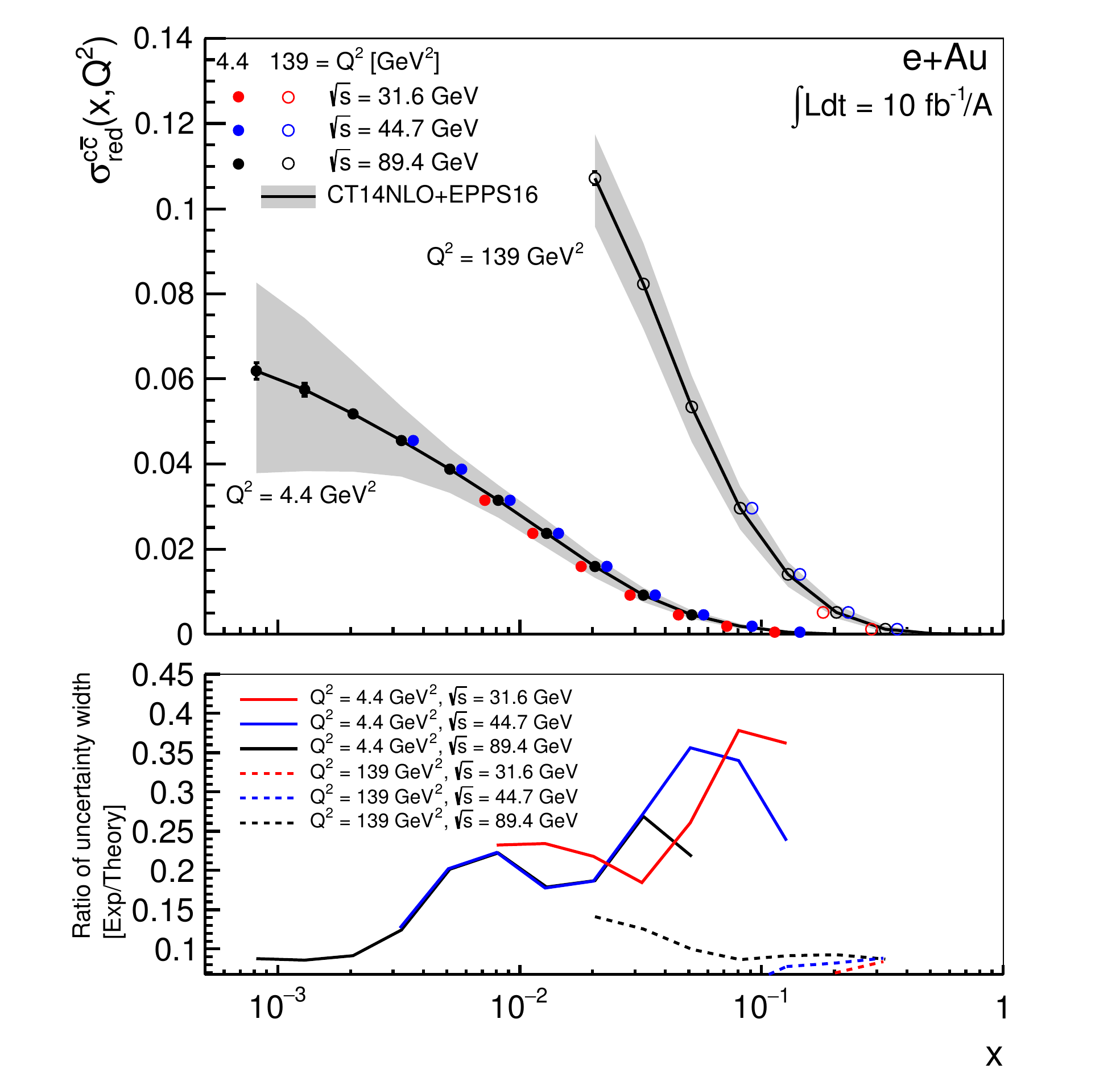}
   \caption{The reduced cross section ({\it left}) of $c\bar{c}$ production in $e$+Au collisions at an EIC is 
plotted as a function of $Q^2$ and $x$. The points are shifted by -log$_{10}(x)/10$ for visibility. 
Two examples of the $\sigma_{r}^{c\bar{c}}$ ({\it right}) at $Q^{2}$ values of 4.4~GeV$^{2}$ and 139~GeV$^{2}$ are plotted versus $x$, with the bottom panel showing the ratio between the widths of the experimental and theoretical uncertainties. Points that correspond to different energy configurations are horizontally offset in $Q^2$ for visibility. The vertical bars represent statistical and systematic uncertainties added in quadrature. The overall 1.4\% systematic uncertainty on the luminosity measurement is not considered in the plot.}
\label{Fig:F2c}
\end{figure}

In the present study, $c\bar{c}$ production events have been selected out of the simulated data sample by tagging $K$ mesons, which are decay products of the $D$ mesons produced in the charm fragmentation. In doing so, we have assumed some $K$ PID technologies to be at place. In particular, an energy loss ($dE/dx$) in the central tracker (i.e. a time-projection chamber) and a proximity focusing Aerogel Ring-Imaging Cherenkov (RICH) detector at mid-rapidity ($-1 < \eta < 1$), covering the $K$ momentum ranges 0.2~GeV~$< p_{K} < 0.8$~GeV and 2~GeV~$< p_{K} < 5$~GeV, respectively. Furthermore we considered a dual radiator RICH detector at forward rapidities ($1 < \eta < 3.5$), covering the kaon momentum range 2~GeV~$< p_{K} < 40$~GeV, and an Aerogel RICH detector at backward rapidities ($-3.5 < \eta < -1$)  covering 2~GeV~$< p_{K} < 15$~GeV.  
In order to suppress the background from DIS events with kaons in the final state not originating from a charm production, we selected only kaons coming from a vertex displaced between 0.01 and 3~cm with respect to the interaction point. 

The overall charm selection efficiency has been estimated to be $\sim30\%$ with no significant c.o.m. energy dependence.
A slight rise with $x$ was also observed, being not significant at very small $Q^{2}$ values and a little more pronounced at higher $Q^{2}$. 

The overall background over signal ratio has been estimated to be respectively 0.95\% ($\sqrt{s} = 31.6$~GeV), 0.98\% ($\sqrt{s} = 44.7$~GeV), and 1.16\%  
($\sqrt{s} = 89.4$~GeV), thus showing a slight c.o.m. energy dependence. This has been also investigated as a function of $x$ at different $Q^{2}$ values for the selected energies and it 
was found to never significantly exceed 2\%.

Figure \ref{Fig:F2c}~({\it left}) shows $\sigma^{c\bar{c}}_{\rm r}$ versus $Q^2$ at different $x$ values, for three c.o.m. energies. 
A log$_{10}(x)/10$ factor is subtracted and points at the same $x$ are shifted for visibility. 
As for the inclusive DIS case, Figure~\ref{Fig:F2c}~({\it right}) also shows two examples of the $\sigma_{r}^{c\bar{c}}$ as a function of $x$ at 
$Q^{2 } = 4.4$ and 139~GeV$^{2}$ without the subtraction of the large log$_{10}(x)$ factor. In the bottom panel, the ratios between the full widths of the experimental and theoretical uncertainties are plotted versus $x$ for the different c.o.m energies. Unlike in the inclusive DIS case, one can see that the experimental uncertainties on $\sigma_{\rm r}^{c\bar{c}}$ at an EIC are expected to be significantly smaller than the current theoretical predictions even at large values of $x$.

\begin{figure}[t] 
   \centering
   \includegraphics[width=0.49\textwidth]{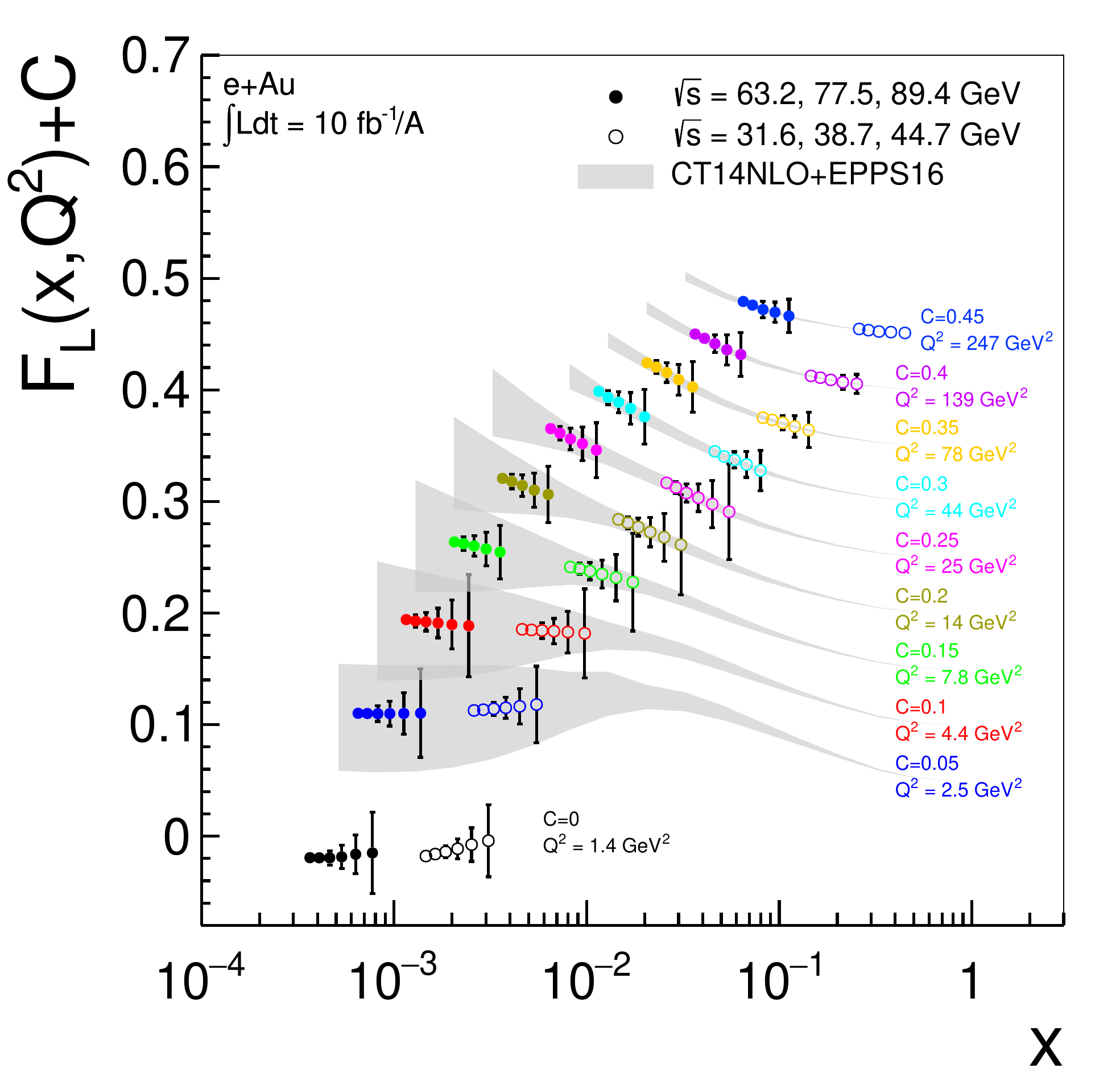}   
   \includegraphics[width=0.49\textwidth]{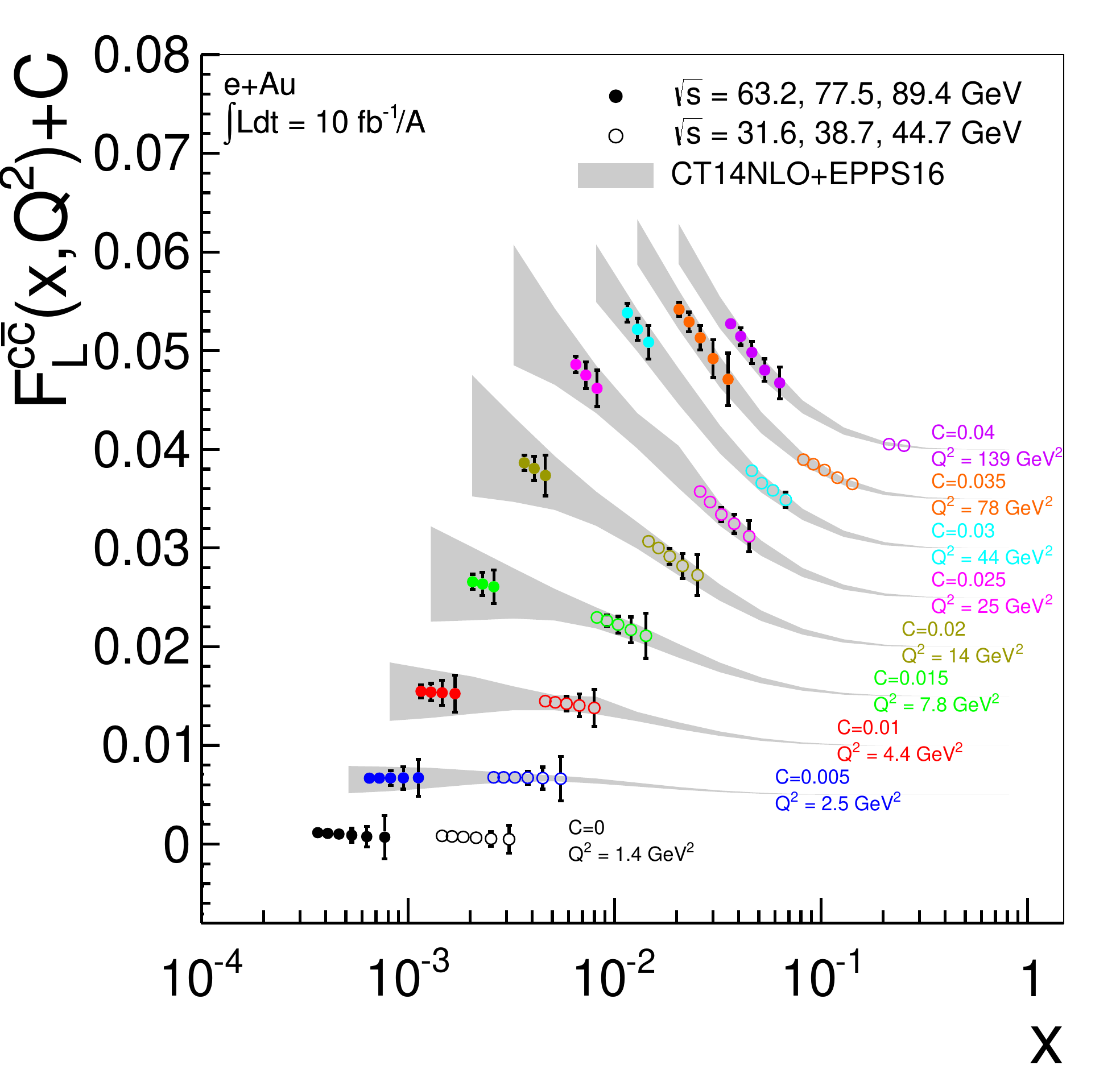}   
   \caption{Inclusive $F_L$ ({\em left}) and $F_L^{c\bar{c}}$ ({\em right}) as a function of $x$ for fixed values of $Q^{2}$, compared to the theoretical predictions based on EPPS16 ({\it gray band}). The vertical bars represent statistical and systematic uncertainties added in quadrature.}
   \label{Fig:FL}
\end{figure}

The extraction of the longitudinal structure function, $F_{\rm L}$, being typically a very small quantity, is experimentally challenging and it is usually achieved through a Rosenbluth separation analysis. This requires measuring $\sigma_{\rm r}$ at different c.o.m. energies. Fitting $\sigma_{r}$ (Eq.~\ref{Eq:SigmaRed}) versus $Y^{+} \equiv y^2/(1+(1-y)^2)$, it is clear that the slope represents $F_{\rm L}$.   
Therefore, having at hand enough range in c.o.m. energy to provide a good lever arm in $Y^{+}$ will be crucial for obtaining good quality fits and a precise extraction of $F_{\rm L}$. 

\begin{figure}[t]
  \centering
  \includegraphics[width=\textwidth]{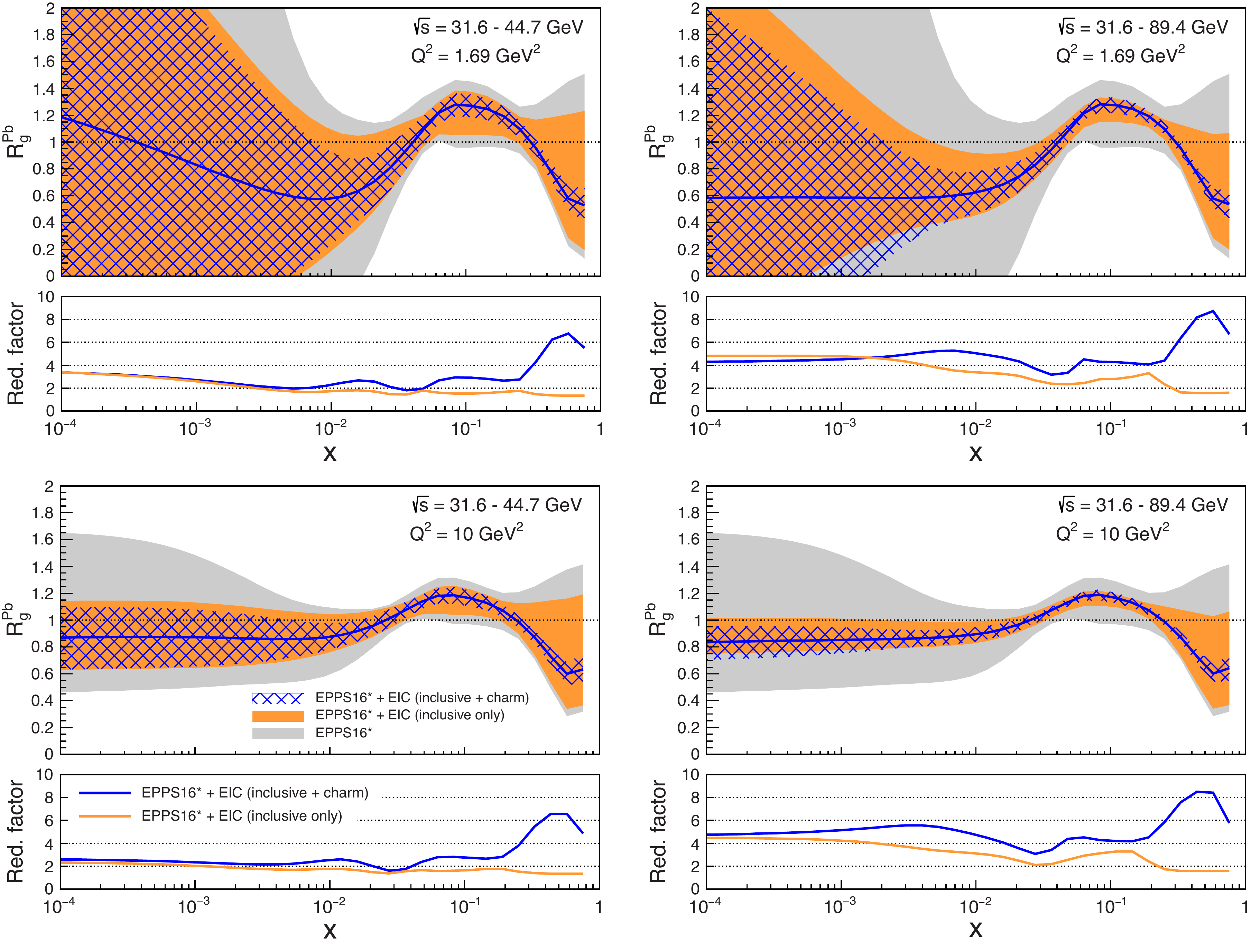} 
  \caption{The ratio $R_g^{\textrm{Pb}}$ of gluon distributions in a $^{208}$Pb nucleus relative to the proton, for lower ({\it left}) and higher ({\it right}) c.o.m. energy, at $Q^{2}=1.69$~GeV$^{2}$ ({upper}) and $Q^{2}=10$~GeV$^{2}$ ({\it lower}). The grey band is the baseline EPPS16* theoretical uncertainty. The orange (blue hatched) band includes also the EIC simulated inclusive (charm quark) reduced cross-section data. The lower panel in each plot shows the reduction factor in the uncertainty with respect to the baseline fit obtained by adding to the latter the simulated EIC inclusive DIS data ({\it orange line}) and also the EIC charm production data ({\it blue line}).}
  \label{Fig:epps16ratios}			
\end{figure}

In the present work we evaluated the potential of an EIC to measure both the inclusive and the charm longitudinal structure functions. Figure~\ref{Fig:FL} shows $F_{\rm L}$ ({\em left}) and 
$F_{\rm L}^{c\bar{c}}$ ({\em right}) plotted versus $x$ for several values of $Q^{2}$. For a clear visualization, the values are offset by adding a constant factor C.
The three c.o.m. energies used in each extraction of $F_{\rm L}$ are also indicated on the plots. Open and solid circles indicate simulations using a 5~GeV and a 20~GeV electron beam respectively.
One can notice that an EIC will be able to perform a very precise measurement of the inclusive $F_{L}$ and $F_{L}^{c\bar{c}}$ in several $x, Q^{2}$ bins, already with a combined collected luminosity of 10~fb$^{-1}$ at each electron beam-energy configuration. 
At its top c.o.m. energies, an EIC can measure $F_{\rm L}$ and $F_{L}^{c\bar{c}}$ with a high precision down to $x \sim 7 \times 10^{-4}$ at low $Q^2$. This is particularly relevant because, at these low values of $x$, the $F_{L}$ predictions by saturation models are already distinctively different from those with a collinear factorization~\cite{Marquet:2017bga}.

In order to estimate the impact of an EIC on the current knowledge of nPDFs, we performed a novel global analysis~\cite{Aschenauer:2017oxs} by adding all the simulated cross section measurements to the currently available experimental data. In this work, to obtain a reliable and less biased estimate of the data constraints, the stiffness of the EPPS16 functional form at small $x$ was partly released by using a more flexible functional form for the gluons, called EPPS16*, and also described in Ref.~\cite{Aschenauer:2017oxs}.

The modification introduced by the nuclear environment, $R_f^A$, $f=q,g$, can be quantified in terms of the ratio between the PDFs of a particular nucleus $A$ and those of a free proton. Figure~\ref{Fig:epps16ratios} shows the resulting nuclear modifications of the gluon distributions caused by a $^{208}$Pb nucleus at $Q^{2} = 1.69$~GeV$^2$ and 10~GeV$^2$, for both lower ({\it left}) and higher ({\it right}) c.o.m. energy configurations.
The EPPS16* theoretical uncertainty from current data only ({\it gray band}) is compared with the result after including in the fit the EIC simulated inclusive DIS data ({\it orange band}) and finally after adding the simulated charm production data ({\it blue hatched area}).
The lower panel of each plot shows the reduction factor in the uncertainty with respect to the baseline fit (gray band). 

Clearly and EIC, especially at it's top c.o.m. energy configuration, has the potential of significantly constrain the knowledge of the nuclear effects on the distribution of gluons over a wide kinematical range. Measuring charm production data, while bringing no additional constraint on the low $x$ region, will have a remarkable impact at high $x$, giving up to a factor 8 reduction in uncertainty. 

We have also studied the impact of the distributions of flavor-separated valence and sea quarks. In spite of some limitation, given by our current analysis not allowing for more flexible parametrizations of 
all the quark flavors simultaneously, our results indicate a significant reduction of the sea quark uncertainties~\cite{Aschenauer:2017oxs}. 

An EIC, with its wide kinematic coverage, flexible ion source and capability of performing high precise measurements of different observables at the same venue, will make possible to perform a stringent test of the universality of nPDFs.

\end{document}